\documentclass[sigconf, natbib=false]{acmart}

\copyrightyear{2026}
\acmYear{2026}
\setcopyright{cc}
\setcctype{by}
\acmConference[CSLAW '26]{Symposium on Computer Science and Law}{March 03--05, 2026}{Berkeley, CA, USA}
\acmBooktitle{Symposium on Computer Science and Law (CSLAW '26), March 03--05, 2026, Berkeley, CA, USA}
\acmPrice{}
\acmDOI{10.1145/3788646.3789520}
\acmISBN{979-8-4007-2447-3/2026/03}

\usepackage[style=trad-abbrv,backend=bibtex]{biblatex}
\usepackage[dvipsnames]{xcolor}
\usepackage{caption}
\usepackage{csquotes}
\usepackage{tikz}
\usetikzlibrary{shapes.geometric, arrows, patterns}
\usetikzlibrary{calc}

\begin{document}

\title{On the Credibility of Deniable Communication in Court}

\author{Jacob Leiken}
\email{jrl9854@nyu.edu}
\orcid{0009-0007-0078-7818}
\author{Sunoo Park}
\email{sunoo.park@nyu.edu}
\affiliation{%
  \institution{New York University}
  \city{New York}
  \state{New York}
  \country{USA}
}

\renewcommand{\shortauthors}{Leiken \& Park}

\begin{abstract}
    Over time, cryptographically deniable systems have come to be associated in computer-science literature with the idea of ``denying'' evidence in court --- specifically, with the ability to convincingly forge evidence in courtroom scenarios, and relatedly, with an inability to authenticate evidence in such contexts. Indeed, in some cryptographic models, the ability to falsify mathematically implies the inability to authenticate.
    Evidentiary processes in courts, however, have been developed over centuries to account for the reality that \emph{evidence has always been forgeable}, and relies on factors outside of cryptographic models to seek the truth ``as well as possible'' while acknowledging that all evidence is imperfect. We argue that deniability does not and need not change this paradigm. 

    Our analysis highlights a gap between technical deniability notions and their application to the real world. There will essentially always be factors outside a cryptographic model that influence perceptions of a message's authenticity, in realistic situations. We propose the broader concept of \textit{credibility} to capture these factors. 
    The credibility of a system is determined by (1) a threshold of quality that a forgery must pass to be ``believable'' as an original communication, which varies based on sociotechnical context and threat model, (2) the ease of creating a forgery that passes this threshold, which is also context- and threat-model-dependent, and (3) default system retention policy and retention settings.
    All three aspects are important for designing secure communication systems for real-world threat models, and some aspects of (2) and (3) may be incorporated directly into technical system design. We hope that our model of credibility will facilitate system design and deployment that addresses threats that are not and cannot be captured by purely technical definitions and existing cryptographic models, and support nuanced discourse on the strengths and limitations of cryptographic guarantees within specific legal and sociotechnical contexts.
\end{abstract}



\maketitle

\section{Introduction}
\label{sec:intro}

Deniable encryption, and cryptographically deniable systems more broadly, refer to privacy-protective communication systems in which parties can convincingly ``deny'' their past communications to a powerful coercive adversary. 
Originally proposed in the 1990s as a theoretical concept~\cite{originaldeniability}, deniable communication has become highly practical since, and reached billions of users through popular messaging applications including WhatsApp and Signal~\cite{signaldocs, collins}.

Over time, cryptographic deniability has become increasingly associated with the idea of ``denying'' evidence in court --- more specifically, of 
(1) 
the \emph{ability to forge evidence} in courtroom scenarios, and relatedly, 
(2) 
an \emph{inability to authenticate evidence} in such contexts.
The association is prominent, with about half of relevant recent papers mentioning courts and/or prosecution as a threat model that deniability addresses.\footnote{Seven of the sixteen papers mentioning cryptographic deniability on the IACR's Cryptology ePrint Archive \cite{eprint} from January 1, 2020 to June 1, 2025 mentioned courts and/or prosecution as a threat model that deniability addresses. The original paper proposing deniability~\cite{originaldeniability} did not mention courts or prosecution.}
However, recent literature has shown that this association has not penetrated legal systems, with courts in the US and Switzerland regularly allowing the introduction of chat transcripts from WhatsApp and Signal~\cite{collins, yadav}.

Mathematical definitions of deniability do directly imply an ability to forge\footnote{In this paper we do not use the word ``forgery'' and its variants to refer to a specific common law crime (as codified, for example, in 18 U.S.C. ch. 25). Rather, we use the term with its colloquial definition: ``The act of forging something, especially the unlawful act of counterfeiting a document or object for the purposes of fraud or deception''~\cite{dictionaryforgery}.} evidence (i.e., (1) above). Within many of the models considered in the deniability literature, if deniability is properly implemented, then there is a mathematical guarantee that any message someone claims to have sent or received is just as likely as any other message to have been the one actually sent or received, even given the cryptographic transcript. Applying this reasoning, treating any particular message as more credible than another would be unfounded (i.e., (2) above).
Indeed, within these mathematical models, (1) the ability to forge messaging history may \emph{logically imply} (2) the inability to authenticate messaging history. 

But in almost any realistic context, the ability to forge messaging history does not imply an inability to authenticate messages, when other contextual and circumstantial evidence and possible witnesses are taken into account. That is, the above logical implication holds only within a mathematical model that strictly limits the information available.

In our view, the deniability literature's focus on courtroom applications is misplaced. 
Even if deniability is implemented perfectly, we expect evidence derived from deniable communication systems to be often relied upon in courts --- not because of a lack of understanding of deniability, but as part of the proper operation of evidentiary systems. 

Courts authenticate evidence using processes that have been developed over centuries to account for the reality that \emph{evidence has always been forgeable}. These processes are heavily reliant on circumstantial evidence and witnesses, and deliberately admit evidence that could conceivably have been forged --- such as verbal testimony, letters, photographs, and so on --- for evaluation by a jury or judge (who may eventually decide they were actually faked). Just like (cryptographically deniable) chat transcripts, these other types of evidence are generally admitted in legal proceedings absent strong evidence of their unreliability, which must go well beyond the mere possibility of forgery. 

Courts were never the killer application for cryptographic deniability, in our view.
That said, deniability is far from useless in court. Deniability may play an important role in limited contexts within legal proceedings, even if the fact of deniability by itself should not be grounds for evidence to be excluded from a courtroom.
In routine use, deniable communication preserves a stronger ability to argue against evidence that lacks strong authentication (i.e., as noted above, most evidence), compared to other more attributable communication methods.
As such, the use of deniable communication places a renewed emphasis on traditional fact-finding processes within the legal system, which were designed to deal with unreliable and forgeable evidence, shifting the focus to questions such as: is the witness who brought the transcript to the court reliable? How was the evidence discovered? Is it corroborated by other parties, and what are their incentives? Is the transcript consistent with the relevant party’s other conduct? 

Morevoer, deniability could be an explicit issue in a legal proceeding if a party raises a challenge to evidence that they can back up with a concrete reason to believe that that particular evidence has been modified or forged --- again, beyond the mere possibility of forgery. This need not be based on technological expertise; it could be as simple as a party to the conversation claiming that they never received the message in question, and perhaps backing this up by pointing to the message history in their app. 
Such a claim would then lead to a more detailed investigation of the evidence, in which technological expertise on deniability might well be relevant. If so, there are existing, if imperfect, processes in the legal system to bring in the testimony of experts.

Just because evidence could be forged does not mean it has been. 
Indeed, if no evidence of a type that could in principle be forged were admissible in court, then pretty much no evidence could be admitted. 
Verbal testimony would be prohibited because people can lie, essentially all documents would be excluded because they can be faked, and items found at the scene of a crime would be inadmissible because they could have been planted.
Even cryptographically signed communications can be forged if the signing key is stolen.

The intended role of evidence law is to ensure accurate fact-finding by, among other mechanisms, filtering out evidence that is clearly irrelevant or prejudicial in certain ways\footnote{For example, a victim's past sexual behavior is inadmissible in cases against their accused abusers. \textsc{Fed. R. Evid.} 412. There are other reasons prejudicial evidence and limited other types of evidence could be excluded, which are beyond the scope of this paper.} --- and to admit the rest, even though it may be unreliable, for the jury or judge to form their own opinion and their own doubts. Although existing evidentiary frameworks are far from perfect, they have been honed over hundreds of years in an effort to administer justice in an environment where almost all evidence is unreliable, the risk of forgery is omnipresent, and these inconvenient facts often present real practical problems (as in the example below).
In this sense, we see deniability as furnishing an additional example of a old and challenging problem with which the legal system is well acquainted, rather than introducing problems that newly disrupt or confound the existing system.

\paragraph{``Deniability'' in court in the Cold War}
The high-profile Cold War case of Alger Hiss, from the 1940s, is one example where the possible forgery of evidence featured prominently. The case elicited extensive commentary on the Anglo-American evidentiary system and the merits and drawbacks of admitting evidence known to be possibly unreliable~\cite{taleoftwo}. Hiss strenuously argued over the course of two trials and an appeal that the government's evidence against him was insufficient. He claimed that they key evidence in the case --- contested letters --- were forgeries written on different typewriters.
Despite contrasting testimony from experts, the letters were introduced as evidence, Hiss's conviction was upheld, and over protracted discussion in the following years, the legal profession decided against making any changes to the system that allowed the contested letters to come into trial~\cite{taleoftwo}. To this day, it is unclear whether the letters were forged. In the absence of certainty, it was determined that admitting the letters was the right thing to do. It allowed the jury to weigh the testimony of Hiss, other witnesses, and experts, to select the verdict they believed reflected the truth. That two juries entered the same verdict, despite different testimony and different presentations by the defense and prosecution, supports the claim that this system works as designed in the face of uncertainty.

\paragraph{Credibility, not deniability}
Acknowledging that there are almost always factors outside of cryptographic models that influence perceptions of an artifact in practice,\footnote{See, e.g., \url{https://www.xkcd.com/538}.}
we introduce a new model of \emph{credibility}, 
comprised of three elements. The first is the \textit{threshold of believability} that must be surpassed for a forgery to be taken as authentic by a relevant viewer in a specific sociotechnical context and threat model; while this is generally not precisely quantifiable, it is nonetheless useful to identify explicitly and reason about. Given this threshold, then, credibility is determined by two related axes: the \textit{ease of forgery} within the system and the default \textit{retention} policy of the system or conversation.

We see credibility as an essential and overlooked component of threat modeling for deniable communications, in realistic contexts not limited to courtroom scenarios. As such, we hope that our conceptual framework can guide the design and implementation of secure communication systems in practice, to facilitate the design of technologies better to equipped deliver on the promise of deniability in diverse sociotechnical contexts.

\paragraph{Outline and contributions}
Section~\ref{sec:related} discusses related work, and Section~\ref{sec:deniability} provides background on deniable communication.
Sections~\ref{sec:credibility} and \ref{sec:evidence} present our novel contributions.
In Section~\ref{sec:credibility},
we introduce the new conceptual framework of \emph{credibility} and explain its implications for deniable communication within and beyond the courtroom.
In Section~\ref{sec:evidence},
we analyze how evidence derived from deniable communication systems is likely to be treated in court, including an overview of existing mechanisms in US legal systems that are meant to deal with unreliable or forged evidence (including complex evidence of a technical nature), and a discussion of these mechanisms' likely application to cases involving deniable communication.\footnote{While our analysis in this section focuses on the US system, many aspects are generalizable. In particular, the general principles underlying rules of evidence and the idea that unreliable or contested evidence may be excluded are reflected in many court systems.}
Section~\ref{sec:discussion} offers discussion and Section~\ref{sec:conclusion} concludes.

\section{Related Work}
\label{sec:related}

In the last two years, four studies conducted by computer scientists aimed to determine legal and/or social perceptions of deniability~\cite{yadav,collins,reitinger,rajendran}. The next two subsections discuss these prior works and how they relate to this paper.

\paragraph{Deniability in Courts}
Yadav et al.~\cite{yadav} and Collins et al.~\cite{collins}, searched judge-written orders and decisions in court cases in the United States and Switzerland respectively for situations where evidence relating to secure messaging was referenced. 
Both papers highlighted that the judges writing the surveyed orders unquestioningly allowed the introduction of evidence from messaging platforms which have cryptographic deniability features. We searched legal databases using the queries developed in Yadav et al.~\cite{yadav} for the time period since the piece's publication and found that the trend has continued. Collins et al. states:

\begin{displayquote}
``Our study, along with with Yadav et al.’s findings show that deniability seems irrelevant in court cases in Switzerland and the United States. Although the results cannot be generalised to other countries, they provide evidence that cryptographic deniability does not translate into real-world deniability and that a purely cryptographic approach does not impact the legal setting. When one party claims a forgery, judges consistently reject these claims.''~\cite{collins}    
\end{displayquote}

Yadav et al.'s discussion section contains paragraphs labeled: ``Deniability is ineffective,'' ``Deniability is hard to achieve in legal cases,'' and ``Even a screenshot is not deniable.''~\cite{yadav} 
These two sets of researchers, with the knowledge and expertise of cryptographers, rightly understand that a message displayed on a platform which has implemented deniability can never be cryptographically verified as legitimate; nor, typically, can a screenshot.
They may moreover be right that not every judge will understand this. From this perspective, it can be surprising or disappointing that judges are willing to permit chat transcripts or screenshots to be presented as evidence, and such an approach may appear correlated to a lack of technical understanding.\footnote{``We need to raise awareness with legal stakeholders that deniability allows bad actors to forge messages and use them in court since courts accept WhatsApp chat~[\textit{sic}] as evidence.''~\cite{yadav} \label{fn:yadav-awareness}}

Furthermore, Yadav et al. write that ``[d]eniability \textit{requires} social and legal acceptance to be effective'' where ``legal acceptance means that the courts accept deniability and may not consider messages as evidence.''~\cite{yadav} 
These statements make explicit a sentiment present in many earlier quotes: that deniability is ``ineffective'' \cite{yadav,collins} in the present state of affairs where judges routinely include evidence from deniable communication systems in court proceedings.

Our paper shares with these prior works the aim of elucidating gaps between cryptographic conceptions of deniability and how courts treat deniable communication in practice, taking into account sociotechnical factors. Our model of \emph{credibility} shares some motivations with Collins et al.'s proposed model of deniability \cite{collins}; but our modeling approaches differ significantly.\footnote{Notably, their ``oracles'' aim to capture a similar concept to our \emph{ease of forgeability}, their variable $\nu$ has a related purpose to our \emph{threshold of believability}, and they have another variable $\mathsf{aux}$ to model contextual information, a concept we also consider important. Their modeling makes more use of cryptographic-style formal modeling techniques (e.g., simulators and oracles) than ours---an approach from which we intentionally diverge.}
However, in contrast to these prior works' view of the current state of affairs as deniability being ``ineffective,'' we argue that the status quo reflects the effective, correct functioning of the evidentiary system, even --- or perhaps all the more so --- given a full understanding of the cryptographic guarantees of deniability. 

\paragraph{Perceptions of Deniability}
Reitinger et al.~\cite{reitinger} and Rajendran et al.~\cite{rajendran} developed prototypes of a Signal-like messaging service with cryptographic deniability built-in. They surveyed potential users for understanding and opinions on the functionality and desirability of such a feature. Reitinger et al.'s experimental setup involved participants who were asked to act as if they were potential jurors in a politician's bribery trial, while Rajendran et al. surveyed participants on whether they would want or utilize this feature in their own messaging. 

These works sought to determine how users perceived deniability, how to best educate people on how and why to use deniability, and to estimate people's preferences, respectively in legal and social contexts. 
Our work has the distinct and complementary motivation of examining how deniable communication is and should be treated under the law of evidence.
The question of how potential jurors are likely to understand electronic communication evidence (with or without deniability), in particular, is closely related to our topics of interest. A better understanding of this question could influence how cases involving this type of evidence are litigated, and how such evidence is presented in courts.

Both studies found that before participants were given explanations about deniability, they were likely to believe chat transcripts presented to them were authentic~\cite{reitinger, rajendran}. However, both found non-cryptographic, non-legalese techniques to get users to question chat transcripts. The simplest technique was to have users interact with a user interface that had implemented deniability in a way intended to enhance conceptual understanding~\cite{reitinger, rajendran}.

These prior works~\cite{reitinger, rajendran,yadav,collins} share a notable focus on user (mis)perceptions of deniable communication; complementarily, our focus is not on perceptions but on legal and sociotechnical analysis based on a complete technical understanding of deniability.

\paragraph{Other Related Works}
A breadth of research across varied disciplines has demonstrated that individuals communicate and understand communication differently depending on the medium and its perceived attributes. Academic marketing literature addressing this topic, for example~\cite{mediumshapesindividuals}, notes that the effectiveness of a message can vary widely by medium because of changes in the perception of the recipient. 
The communication technology in use changes the way individuals communicate to make decisions, making a group more or less collaborative~\cite{mediummoderatesgroups}. 
Intuitively, users are less likely to share information that can be categorized as ``intimate'' if they are communicating over a channel that they perceive to be open or insecure~\cite{mediumcomfort}. Individuals tend to change their behavior when they believe they are being surveilled, whether or not they are in fact~\cite{eyesbehavior}, and tend to speak less freely when they know their communication can be monitored~\cite{trustnotenough}.

Many privacy theorists struggle with a ``privacy paradox,'' that individuals often choose to share more information about themselves publicly as communications technology has advanced, a decision in apparent conflict with their stated value for privacy~\cite{AG05,knowncitizen}. Some prominent theories of privacy suggest that at least some of these apparent conflicts may arise from conflating different notions of privacy~\cite{privacy-paradox-myth,contextualintegrity}. Other literature has shown that users tend to be unaware of data practices and don't often know what data is being collected about them and for what use~\cite{vankleek2017}, and that it is onerous or impractical to maintain such detailed knowledge~\cite{nyt-privacy-policies,MC09}. Voluntary disclosure aside, individuals still view a choice to disclose information differently from (risk of) exposure that is inadvertent or against their will~\cite{knowncitizen}.

The above literature shows that individuals communicate and understand communications differently based on perceived security and privacy properties of the communication medium. A diversity of options allows individuals to choose the technology most suited to purpose, and deniability occupies an important space in that spectrum.

\section{Deniability}
\label{sec:deniability}

\paragraph{What is deniability?} 
The simplest meaning of ``deniable'' is ``possible to deny,'' and the verb ``deny'' is defined commonly as to declare untrue or refuse to believe or acknowledge (e.g.,~\cite{dictionarydeny}). 
The dictionary definition we found most relevant to the context of deniable communication within computer science is ``being such that plausible disavowal or disclaimer is possible''~\cite{dictionarydeny}. Technical definitions within computer science consider the disavowal or disclaimer of either the content of a communication (\emph{transcript deniability}),\footnote{As in \emph{deniable encryption}, e.g.,~\cite{originaldeniability}.} or its authorship or provenance (\emph{conversation deniability}),\footnote{As in \emph{deniable authentication}, e.g.,~\cite{deniableauthentication}.} often by providing plausible evidence that a different communication took place instead. Broadly, the goal of deniable communication systems is to eliminate or reduce the availability of cryptographic evidence that would provide convincing documentation of the contents or provenance of an electronic communication.
As further discussed below, the precise technical definitions proposed to achieve this goal vary, as do the adversaries and threat models they address.

\paragraph{A walk in the woods}
A commonly invoked scenario to illustrate the ideal of deniable communication is a ``walk in the woods.'' A conversation held during a walk in the woods is ephemeral, as is the mere fact that the conversation occurred. As such, the contents of the conversation can only be repeated through the individuals' memory. Whether or not a third party believes that the conversation occurred, or the relayed contents, depends in practice on their view of the trustworthiness of the person repeating the conversation.\footnote{A party could surreptitiously record the conversation, and this has gotten progressively easier with modern recording technology. But importantly, the ``default'' way that a walk-in-the-woods conversation happens is without surreptitious recording. That is, a party must go out of their way in order to make and conceal a recording, and moreover must do so in the moment as it becomes impossible after the fact. In this paper, when we discuss the walk-in-the-woods ideal, we generally assume there is no surreptitious recording.}

\medskip

In contrast, electronic communication often creates long-lived evidence trails distributed across multiple parties including non-parties to the communication (such as intermediary platforms), which can systematically support convincing and detailed documentation of past communications.\footnote{Such evidence is not irrefutable, as we discuss more later. For example, an adversary who compromised a user's password could send messages in the user's name that would pass cryptographic verification, but would in fact not have been sent by the user. Our point here is that such evidence trails are quite reliable in general, and available systemically.}
This can be the case even for communication systems that support strong privacy guarantees of other types, such as end-to-end encrypted (E2EE) systems. 
This is not an oversight, but rather, simply not the type of threat that E2EE is designed to protect against. 

A key motivation of deniable communication is to create electronic communication systems that have similar deniability properties to the walk in the woods. Why? Secure and private communication, including deniable communication, is valuable for privacy and freedom of expression.\footnote{For example, research has shown that people speak less freely when they know their communication can be monitored~\cite{trustnotenough}, and the UN Office of the High Commissioner on Human Rights has emphasized that strong encryption is essential to fundamental human liberties including free expression~\cite{ohchr}.} Deniable communication techniques can also be valuable as a building block for other security- or privacy-enhancing systems.\footnote{The original proposal of deniable encryption suggested two fairly specific cryptographic applications: receipt-free voting protocols and incoercible multi-party computation~\cite{originaldeniability}.} Furthermore, research on deniable communication can enhance our understanding of the theoretical and practical boundaries of cryptographic techniques. It is debatable whether an electronic communication system could ever achieve deniability commensurate to in-person communication, and it is an interesting open question how close we can get both in theory and in practice.

\paragraph{Deniable communication primer}
Standard secure communication often involves two components: (1) \emph{authentication}, where senders attest that they are the originators of the messages, in a way that recipients can verify, and (2) \emph{encryption}, where senders encrypt content before sending, in such a way that only intended recipients can decrypt it. Both these properties rely on the secrecy of some information known only the sender or recipient, respectively, called a \emph{secret key} (much like a password). By using their secret key, a sender can attest that they wrote a message by \emph{cryptographically signing} it, and they can decrypt messages addressed to them. Nobody else, without the relevant secret key, has these capabilities. But if an adversary compromises a secret key, then all bets may be off: the adversary may be able to impersonate and decrypt messages intended for the victim. A cryptographic transcript and the associated secret keys would typically constitute pretty solid evidence both of who participated in a conversation and what they said.\footnote{Someone who had both of these could use the keys to decrypt the transcript themselves --- and typically, without deniability features designed specially for the purpose, it is not feasible to decrypt encrypted transcripts in multiple plausible-looking ways.}

Deniable encryption, then, adds a layer of security on top of standard encryption, to provide protection even against a strong coercive adversary who can demand access cryptographic transcripts and secret keys from conversation participants, by allowing the participants to generate alternative secret keys that look consistent with the cryptographic transcript of the conversation, but make it look like either (1) something else was said (\emph{transcript deniability}, which relates to encryption) or (2) they did not participate in the conversation (\emph{conversation deniability}, which relates to authentication).
For example,  if Alice and Bob are using an encryption scheme which has implemented cryptographic deniability, the result of Eve's ``decryption'' could be any string of Bob's choosing, or a decoy string set in advance.

Cryptographically deniable systems are already in widespread use. Most notably, the Signal protocol, used by the popular end-to-end encrypted WhatsApp and Signal messaging apps as well as other secure messaging systems, implements conversation deniability~\cite{vatandas}. This allows conversation participants to deny their participation (conversation deniability), but not to deny the contents of specific messages (transcript deniability).

If two people are communicating using an encryption scheme that implements (either kind of) cryptographic deniability, this precludes any cryptographic evidence that a specific message has been sent by a specific person~\cite{originaldeniability}, \emph{either} because the conversation \emph{or} the transcript is deniable. If Bob ever leaks a message from Alice, Alice has a mathematical way of repudiating that message. The mere possibility that any participant can repudiate a message in this way shows that Bob \textit{could} be lying. Thus, conversations over deniable encryption bring us a step closer to the ``walk in the woods.'' With message deniability, for example, there may be evidence that a conversation occurred (Eve could have watched Alice and Bob walk into the trees), but proving what was said has become much harder. Eve may eventually think that she knows what was discussed and who said what, but Eve's impression would have to be based on interviews with Alice and Bob and her perception of their trustworthiness.

\paragraph{Variants of deniability}
A range of variant definitions of deniability exist. For this work, we construe the term broadly. As our work is largely agnostic to definitional variants, we refer to~\cite[\S1.5]{fullydeniable} for an overview of variation in cryptographic definitions. We highlight one non-cryptographic deniability notion, as it relates to our later discussion of credibility:
\emph{hidden volumes} constitute a form of deniability in the context of operating systems. VeraCrypt, a fork of the now-defunct application TrueCrypt, provides a fairly mainstream implementation of this~\cite{falsebottom}.\footnote{TrueCrypt is said to have been downloaded 30 million times over the 10 years it was active~\cite{truecryptusage}.} 
It allowed users to type a preset fake password into a locked computer, revealing a ``safe'' partition and hiding the original~\cite{compelleddecryption}. This allowed users to deny the existence of files saved on their device. In this case, a user acknowledged that a device was their own, but effectively denied the existence of whatever was on the hidden partition.\footnote{This is not a true implementation of deniability. Rather, it is referred to as ``false bottom'' encryption. As far as we are aware, true deniability has never been implemented for data at rest. Deniability for data at rest is generally outside the scope of this paper.}

\section{Credibility: Bridging Cryptographic Deniability and Real-World Attribution}
\label{sec:credibility}

Cryptographic deniability is an important property with precise mathematical definitions. But as discussed above, for reasons that lie beyond the scope of cryptographic models, even perfectly implemented strong deniability does not render communications impossible to attribute in practice.

We propose a new framework to address this gap. Our new definition of \emph{credibility} breaks down the key sociotechnical and contextual considerations that influence the imperfect art of attribution in real-world contexts. Cryptographic deniability and other technical system properties are important factors that influence credibility; yet, credibility is an inherently sociotechnical concept, and cannot be analyzed based on technical properties alone.

Analyzing communication systems through the lens of credibility makes explicit the sociotechnical considerations that impact how ``denials'' are enacted and received in practice, providing a much-needed language to reason about critical aspects and likely outcomes of deniable system use, which are typically considered ``outside the model.'' 

Next, Section~\ref{sec:credibility:3gaps} elaborates on the gaps we seek to bridge, Section~\ref{sec:credibility:3elements} presents our credibility framework and its three elements, and Section~\ref{sec:credibility-discussion} provides broader discussion on our framework.

\subsection{Three gaps to bridge}
\label{sec:credibility:3gaps}

\paragraph{Gap \#1: Is the record preserved?}
Written (or otherwise recorded) communication is inherently less repudiable than an oral conversation due to the creation of the record.
Consider the example of a handwritten letter. Anyone (with a writing sample, time, and patience) could produce the exact same letter, yet in many realistic contexts, the document would be considered stronger evidence than an oral retelling.

Deniable communications schemes are likewise less repudiable than a walk in the woods. If Bob shows Eve a chat transcript on his phone of a conversation between Bob and Alice, there may be no proof that Alice actually sent the displayed messages, but the display in front of Eve may well give her more confidence in their veracity and detailed accuracy than Bob orally relaying a conversation. This intuition reflects the work required by Bob in both states. If the transcript were real, then there would be a record that looks exactly like this which would be easy for Bob to show Eve. If it were fake, Bob would have had to undertake some effort to create the fake, thereby diverging from the default condition of the system. This state of affairs contrasts with a walk in the woods, where the work required to relay the true or forged contents of the conversation is effectively identical.

Similarly, if Alice and Bob communicated over a platform implementing cryptographic deniability and Eve is a cryptographer who is in a position to examine Bob's device memory, then if the secret keys and randomness are on Bob's device at the time of Eve's (surprise) examination, Eve may be more likely to believe them as the originals. Although Bob could in principle have prepared fake secret keys and randomness and left them on his phone for Eve to find, this is not the default state of affairs; Bob would have had to go out of his way to prepare the fakes. As such, Eve may or may not believe it is more likely that the secret keys and randomness are unmodified, depending on the context.

In all these cases, the starting point is the original records proffered, records which remain intact unless someone deletes them or goes out of their way to do tampering or forgery: the letter, the message, or the cryptographic material.
The field of \emph{forensic science} encompasses scientific approaches to investigating and analyzing such records and their traces which may be used as evidence in legal proceedings (e.g., \cite{nist-forensics,aafs-forensics}). Broadly, forensic science concerns all kinds of evidence, from handwritten letters to blood splatters and much more. Digital forensics is the sub-field that focuses on digital evidence and investigations (e.g., \cite{forensics-textbook1,forensics-textbook2}). 
Successfully passing off a forged letter or a forged Signal transcript as authentic is much more difficult under forensic examination than towards casual observers; and where a plausible doubt is raised about a particular piece of evidence in court, forensics experts are likely to be called in.\footnote{While this is not a very high bar, we note again that the threshold is to cast concrete doubt on the authenticity of a specific piece of evidence (e.g., this particular letter), and pointing out the possibility of forgery of evidence (e.g., letters in general) in a given medium, without more, is generally not enough.}

The walk in the woods mitigates this issue by minimizing the possibility of any original records of the communication being available at all. Electronic communication media generally leave more traces than a literal walk in the woods; nonetheless, in a similar spirit, we can design systems that minimize retention of records. 
A common example available on many popular messaging platforms is ``disappearing messages.''\footnote{``Disappearing messages'' aren't perfect in practice, but they shift defaults of retention in an impactful way~\cite{aclu-disappearing}.} A more thorough alternative might involve constantly wiping keys and refreshing randomness in device memory. Indeed, the Signal protocol does something similar~\cite{signaldocs};\footnote{Signal deletes per-message keys upon decryption.} that said, simple deletion in software is unlikely to withstand forensic examination in practice, and secure deletion is very difficult to get right~\cite{insecurekeys,fairoze-thesis}.\footnote{See, for example, \textit{E.I. du Pont de Nemours \& Co. v. Kolon Indus., Inc.}, 803 F. Supp. 2d 469 (E.D. Va. 2011), sanctioning a defendant company for deleting files after the start of litigation. Although the deleted files were unrecoverable, the deletion itself was proven by a forensic analysis of recovered metadata.} 
These types of features are not cryptographic in nature,\footnote{It is not currently possible to prove non-retention cryptographically. Recent papers have suggested quantum techniques to do so, but these are not practically feasible at the moment~\cite{quantum-non-retention}.} so naturally lie outside cryptographic deniability definitions; they relate to system design and defaults, and are still within the scope of security engineering.

\paragraph{Gap \#2: How much effort?}
The preceding hypotheticals note not only that Bob would have to go out of his way to forge records, but also the effort that Bob would have had to invest in so doing. Existing definitions tend to focus on whether or not forgery is possible, and treat forgery as ``easy'' if there exists an efficient algorithm to perform it. In practice, the perceived likelihood of forgery may depend on how much effort is required. While the concrete efficiency of the algorithm is one relevant consideration, many other practical considerations are non-technical or only partly technical in nature. Do forgery tools come with the communication platform, or would they have to be obtained separately? Are there existing libraries or applications that perform forgery, or would one have to write the code oneself? Does forgery require technical expertise? Are existing tools easy to use? For a coder, or for a layperson? Could a layperson pay an expert to do the forgery instead? How much would that cost, and how hard would it be for them to find a competent and discreet expert?

\paragraph{Gap \#3: Who is trying to convince whom, in what context?}
Finally, context and relationships matter in whether someone is likely to be believed. 
For example, if Eve knows Bob is a serial liar who has faked letters before, she will be less likely to believe him (even if provided a detailed communication transcript). If Bob is a long trusted friend or family member of Eve, then she will be more likely to believe him (even without a detailed communication transcript). In either case, if Alice is also friends with Eve and corroborates Bob's story to Eve directly, this may boost Bob's credibility; if Alice says Bob is lying, this may damage his credibility. Incentives and content are critical contextual factors. If Alice is Bob's creditor and she corroborates that she released him from a debt, that is especially credible. 
A chat transcript between politicians about matters of national concern, or a chat transcript that appears to be evidence of a crime, will receive much more scrutiny than personal messages of no public concern that do not make anyone look bad.

\medskip

The three gaps that we identify highlight a range of contextual ``outside-the-model'' factors that feature in realistic assessments of credibility --- whether or not cryptographic attribution methods are available. The importance of contextual factors in assessing credibility is naturally heightened in those situations where cryptographic attribution is not available --- that is, most situations, as most kinds of evidence have no cryptographic attribution. 

Within a cryptographic model, if messages can in principle be forged, there may be nothing more that can be done to ascertain the likelihood that a given message has been forged --- that is, forgeability may imply impossibility of attribution.
But outside of the cryptographic model, as underscored by the discussion above, it is essentially never the case that there is nothing more that can be done to ascertain the likelihood that a specific message in a specific context has been forged. Neither Eve, nor the judge in the courtroom scenario, needs to pack up and go home upon discovering the possibility of forgery; instead, they turn outside the cryptographic model, to contextual factors that they would have considered in any case, but are now their primary recourse. And indeed, the contextual factors highlighted above are the sort of things that could well feature in courtroom scenarios trying to assess the reliability of evidence.

\subsection{The three elements of credibility}
\label{sec:credibility:3elements}

Crystallizing the above discussion, communications technologies can be seen as ``implementing'' credibility on two axes. 

The first is message \textbf{retention} (related to Gap \#1). The less time a message is retained by default, the more often it will be credible for message participants to say that they no longer have the original. As discussed in Section~\ref{sec:credibility:3gaps}, retention is outside the scope of cryptographic deniability.

The second is \textbf{ease of forgery} (related to Gap \#2). If it is easy to create a false message that will be believable as a possible original message, then the mechanism implementing deniability also benefits credibility. Cryptographic deniability focuses entirely on this axis. As noted in Section~\ref{sec:credibility:3gaps}, technical features such as cryptographic deniability are necessary but not sufficient to determine ease of forgery in practice, as other contextual factors also matter. 

The third factor of our credibility framework is the \textbf{threshold of believability} (related to Gap \#3). That is, the bar which a forgery must pass to be likely to be believed in a relevant sociotechnical context. As we explain further below, unlike the preceding two axes, the threshold of believability is independent of the communication medium.

The following subsections provide more detailed discussions on the three elements and how they fit together. 

\subsubsection{Retention}
\label{sec:credibility:3elements:retention}

The retention of past communications refers to the default amount of time a message is retained after viewing.
This default could be either at the service level or the conversation level.
Any communication could be recorded without consent, such as a surreptitious audio recording of a conversation during a walk in the woods or a screenshot of a message sent in Snapchat. However, these require additional effort in the moment, meaning that parties to the communication would have to be actively adversarial at the time of the communication. This axis instead measures the default where the parties communicating are non-adversarial. Retention varies greatly, with one end being no retention past the initial viewing (i.e. a walk in the woods, Snapchat), and the other being permanently retained by many parties (i.e. an official government edict). 

Disappearing messages are sometimes not included in discussions of deniability because they are not cryptographic. 
But retention is an important aspect of secure system design, in part because disappearing messages can mirror many of the characteristics of the prototypical walk in the woods. Assuming messages are ``disappeared'' thoroughly, which comes with its own host of technical challenges, a disappearing message also means that there is, by default, no record of a conversation after it happens. The Signal protocol implements deniability through a short-retention scheme~\cite{signaldocs}, so significant thought has gone into this axis in practice, albeit not in this explicit framing.

\subsubsection{Ease of forgery}
\label{sec:credibility:3elements:ease}

\begin{figure*}[ht]
    \centering
    \begin{tikzpicture}
    
        \draw[help lines, color=gray!30, dashed] (-4.9,-4.9) grid (4.9,4.9);
        \draw[->,ultra thick] (-5,-5)--(5,-5) node[right]{\textbf{Ease of Forgery}};
        \draw[->,ultra thick] (-5,-5)--(-5,5) node[above]{\textbf{Retention}};
        
        \filldraw[black] (4.8,-4.8) circle (2pt) node[anchor=east]{A walk in the woods};
        \filldraw[black] (3,2) circle (2pt) node[anchor=west]{Experimental Signal variant in~\cite{rajendran}};
        \filldraw[black] (-2,2) circle (2pt) node[anchor=west, align=left]{Signal as implemented \\ in August 2025};
        \filldraw[black] (0,0) circle (2pt) node[anchor=east, align=right]{A photo or video purportedly \\ depicting a real person or events};
        \filldraw[black] (1.5,0) circle (2pt) node[anchor=west]{A screenshot of Snapchat};
        \filldraw[black] (-4.8,4.6) circle (2pt) node[anchor=west]{Statutes, news articles, misc. independently verifiable records};
        \filldraw[black] (-4.8,-4.8) circle (2pt) node[anchor=west]{Citywide tornado siren};
        \filldraw[black] (-4,4) circle (2pt) node[anchor=west]{DKIM-verified email};
        \filldraw[black] (4,3) circle (2pt) node[anchor=west]{Unsigned letter};
    
    \end{tikzpicture}
    \captionof{figure}{Different communications techniques on the two axes of credibility}
    \Description[A plot placing various communications techniques on the two axes of credibility]{Ease of forgery is the x axis. Retention is the y axis. A walk in the woods is at (4.8,-4.8), experimental Signal variant in~\cite{rajendran} is at (3,2), signal as implemented in August 2025 is at (-2,2), a photo or video purportedly depicting a real person or events (0,0), a screenshot of Snapchat is at (1.5, 0), statutes, news articles, misc. independently verifiable records is at (-4.8,4.6), citywide tornado siren is at (-4.8,-4.8), DKIM-verified email is at (-4,4), unsigned letter is at (4,3).}
    \label{fig:securityplot}
\end{figure*}
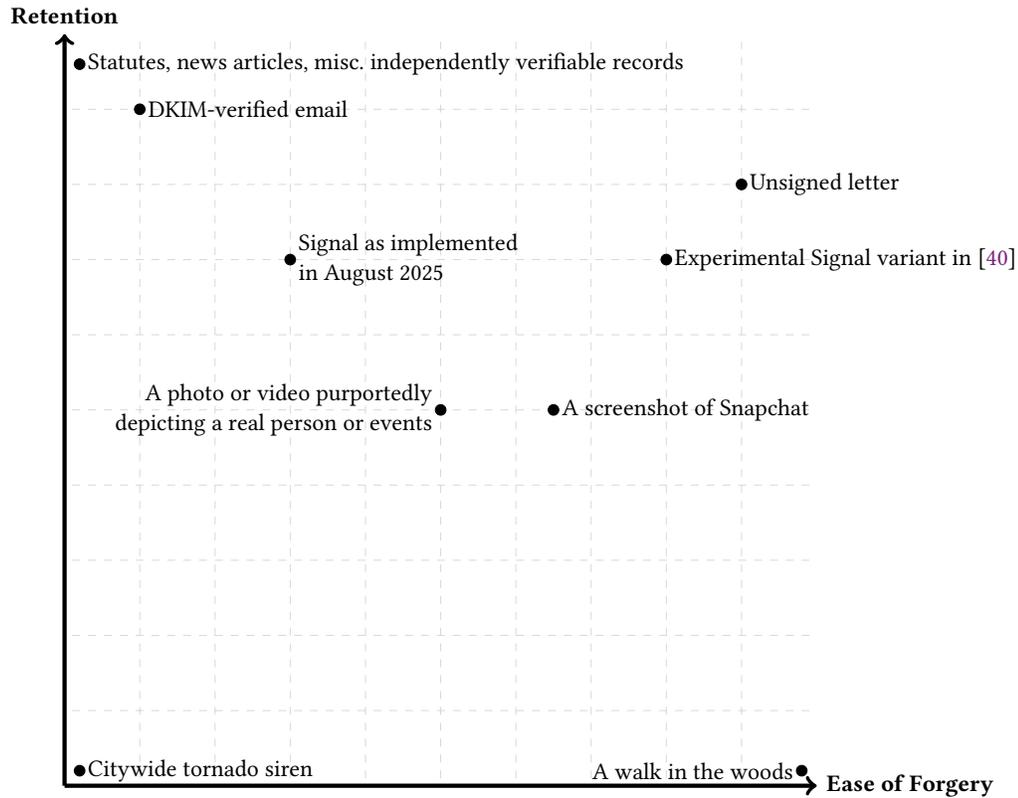

A forgery is relatively ``easy'' if it can be easily done by an untrained individual using readily available resources or cheaply done by a trained individual --- either means that it is widely accessible.\footnote{However, paying a trained individual creates an additional potential vulnerability outside the cryptographic model, in that the trained individual now knows about the forgery and could be a potential witness.}\textsuperscript{,} \footnote{Often, as the technical skill becomes more specialized, the cost of performing the skill as a service increases. But this is not true of deepfakes created using generative AI, which are increasingly used to create images passing a requisite threshold of believability. The technical skill required to make a convincing deepfake is fairly high, but skilled individuals currently offer to create deepfakes for relatively cheap because online platforms, at least until recently, have incentivized this behavior~\cite{mrdeepfakes}.} 
One end of this axis is deniability by nature or forgery that is as easy as telling a lie (e.g., a walk in the woods), 
and the other is cryptographically non-repudiable messaging with no way to edit messages (e.g., an encrypted, cryptographically signed email).\footnote{The relevant threshold of believability may affect where on the ease of forgery axis a service falls. Consider a technology which allows for the easy creation of non-believable forgeries but which requires exponentially higher skill and time to make believable forgeries. For a low threshold, this service would be very high on this axis, but for a high threshold, it would likely be lower. Having one plot, as in Figure~\ref{fig:securityplot}, assumes that the rate at which the quality of the forgery improves is proportional for all technologies depicted.}
In between these extremes, towards the ``easy'' end, we place deniability implemented through an easy-to-access, easy-to-use interface built in to a communication platform (e.g., as suggested in \cite{collins} and prototyped in~\cite{rajendran}).\footnote{Notably, existing applications do not currently offer a readily accessible, user-friendly interface for denying communications, even where they do have cryptographic deniability features. The example in \cite{rajendran} was a research prototype only.}
While the computational efficiency of the forgery process is a relevant factor, we find that socio-technical and non-technical considerations tend to dominate the analysis of ease of forgery in practice (e.g., in all examples in Figure~\ref{fig:securityplot}).

\subsubsection{Threshold of believability}
Consider the threshold for creating a credible forged photograph of Alice's head on Carol's body. Say the photograph looks believable at first glance but when zoomed in there are pixels that are clearly artifacts of an edit in Photoshop. For a personal post to a small audience on Instagram, this might pass the threshold of believability. But if that Instagram post went viral and made the national news, then the heightened scrutiny it would receive would likely bring these artifacts to light. Similarly, for a law enforcement investigation or court case, this forgery likely would not pass the threshold. In other words, a forgery that suffices for one purpose may not for another. This simple fact is made significant by the reality that there is almost always scope to consider contextual factors more thoroughly in assessing an artifact's credibility --- so how thoroughly people are likely to investigate contextual factors ends up a critical consideration for credibility in practice. Court cases, and especially criminal cases, present a context which is specifically designed to try to bring into consideration as many contextual factors as feasible, even at relatively high cost, as someone's life or liberty may be at stake.

We define the threshold of believability as independent of the communication medium; instead, it focuses on questions such as who is trying to convince whom, and in what context, as discussed under Gap \#3 in Section~\ref{sec:credibility:3gaps}.
Although we have introduced it as our third element of credibility, it will often be the first point of inquiry in analyzing the credibility properties of a real-world application context, such as when designing systems. Deciding on a(n approximate) threshold of believability as a initial step allows a more structured inquiry into the two axes with respect to the chosen threshold. The relevant threshold of believability in a given context could help system designers determine where on the other two axes of \emph{retention} and \emph{ease of forgery} a service should be located to provide the desired level of credibility for an application context, or for individuals deciding which communications technique to use. In this sense, the threshold of believability may shade an area on the graph constructed by the two axes as a target; it is not a third axis in itself.

\subsubsection{Big picture}

These three elements are distinct but overlapping ways of approaching credibility, not fully separate from each other. No direction or quadrant on the axes is ``better'' than another; the best configuration for an application depends on its context. Identifying methods of communication as existing in this problem space is helpful both to better understand the space of possibilities, and to promote a diversity of options.

In Figure~\ref{fig:securityplot}, we plot a variety of communication methods on the first two axes of credibility. We have opted for a two-dimensional plot as the third element, the threshold of believability, is independent of the communication medium as noted above. None of the elements are precisely quantifiable or defined by a precise metric, due to inherent qualitative and socio-technical considerations, so there is necessarily some imprecision involved in the plot. We believe the graphical representation is nonetheless helpful to convey our own approximate and comparative sense of where different communication technologies lie within the new space that our credibility framework defines.

\subsection{Discussion on credibility}
\label{sec:credibility-discussion}

\paragraph{How does credibility connect deniability to real-world attribution?}
There are many different reasons credibility can be negatively affected, including easy external verification --- someone may claim that the front page of today's New York Times says $X$, but it is generally easy for a listener to determine the veracity of that claim, so the claim is not credible if New York Times front page in fact does not say $X$. For rather different reasons, a tornado siren would be difficult to fake credibly, because it relies on city-, or even region-wide equipment, and should be perceived by an entire region of people at once if it in fact happened. (Both of these examples appear in the plot in Figure~\ref{fig:securityplot}.)

Examples such as these illustrate that credibility expands on the scope of deniability in a way that captures important aspects of whether true and false claims about past communications are likely to be believed in practice, which are outside of the scope of deniability alone. As theorized in literature, deniability is intended to eliminate the possibility of the coercer ever believing they have the true plaintext, thereby making the whole exercise of coercion futile~(e.g.,~\cite{originaldeniability}). But the mere fact of the ability to generate false secrets realistically carries little significance if the coercing party believes, for reasons outside the model, that they have a reasonable likelihood of having decrypted the true plaintext. Thus, credibility is the practical stopgap connecting purely technical notions of deniability to real-world attribution. 

Put differently, deniable encryption (resp., authentication) ensures that Bob, in assessing claims made about a plaintext (resp., Alice's participation) in a disputed interaction, cannot credibly point solely to the claimed original record of communication, but rather must rely on contextual information beyond that record --- and that is where credibility comes in. Deniability does not imply an inability to attribute, but rather, pushes us to look beyond the transcript by eliminating the apparent ``smoking gun'' of the link between the message, its encryption, and claimed keys.\footnote{As also noted above, even this link is not a true smoking gun, as keys may be compromised.}

\paragraph{Why does credibility matter?}
As previously noted, no point in Figure~\ref{fig:securityplot} is the ideal; we define this concept to highlight the importance of the availability of options that mimic the full range of analogue communication. 

There are many situations where a difficult to forge, permanently retained communication is the ideal. For example, the buyer of a house would be unlikely to go through with the transaction unless they were promised a difficult to forge, permanently retained contract. On the other hand, young adults developing their self expression may value or benefit from low retention, and from scenario to scenario, prefer more or less credible forgeability~\cite{teenagersinternet}.
The related literature discussed in Section~\ref{sec:related} suggests that developing a variety cryptographic techniques enabling digital communications services to be provided at many points on the axes of credibility has a positive impact on the lives and welfare of everyday individuals, whether or not they are aware of the advances. Having more variety in the credibility properties offered by digital communications technologies, including the ability to mimic a wider spectrum of analog communication methods, allows individuals freer choice over proper technology for their expression.

These considerations are outside the cryptographic model, but it is still important to have a structured way of approaching them. Credibility provides such a framework for designing systems that take into account aspects of real-world threat models that would previously have been considered outside the scope of deniability.

\paragraph{The three elements of credibility are not comprehensive, but we believe they capture essential aspects.}
While no theoretical framework can include all practical considerations, we believe that the three elements we identified present a valuable approach to reasoning about credibility, which capture aspects beyond those explicitly named in the discussion above. 

For example, usability is an important consideration not directly not captured by the three elements, but it interacts in subtle ways with all three. If deniability is implemented in a way that users do not understand, they may not use that feature or may use it incorrectly, and would therefore lose out on the security protections it promises. 
A system with poor usability may make creation of forgeries more difficult, which would be reflected under ease of forgery.

We hope that elaborating our theory of credibility, and deeper analyses of how exactly real-world cryptographic deniability technologies (with diverse parameter configurations) fit and relate to one another within the credibility space we have introduced, can be a fruitful direction for future inquiry.

\section{How Evidence Is Authenticated in Court}
\label{sec:evidence}

Courts have developed a variety of mechanisms to determine what evidence can be presented at trial, within the longstanding context that most evidence introduced in courts does not have the possibility of cryptographic verification. In the US system, judges decide whether evidence can be admitted and the evidence that is admitted is presented to the jury, who may choose to take it into account in rendering a verdict. The jury does not need to view the evidence presented as inherently reflecting the truth of the matter; it is instead presented to help the jury determine what the truth of the matter is, on the understanding that the jury should do their best to determine the truth even though some of evidence presented may be contradictory or unreliable. 

Consider the example of verbal testimony, which is always forgeable in principle, as people can lie. What is more, ``he said, she said'' situations do arise in courtroom scenarios, and in general, both witnesses are permitted to testify --- even if, logically speaking, at least one of them must be lying. It is the jury's role in the system to consider the accounts presented to them and use their best judgment to reach a verdict. To try to ascertain who is right and who is wrong, and only present the one who is right, would create a recursive problem --- it is the jury's role to consider all the evidence and determine who is right or wrong, and taking that away from them not only invalidates their role in the system (and the defendant's right to a jury trial, where applicable), but also leaves wide open the question of how then to reach such a determination, having thrown out the very mechanism designated by the legal system to do so. To block both stories from being presented, on the grounds that they are possibly unreliable --- or to exclude all verbal testimony on the grounds that it is, in principle, forgeable --- could be to block what might be some of the most relevant information available to the jury, and leave them little or nothing to work with. On balance, legal systems have determined that admitting some possibly unreliable evidence, and letting the court system take it all into consideration with the knowledge that it may be unreliable, is ultimately more conducive to truth-seeking than the alternative.

At the same time, legal systems do have checks and limits on what evidence can be admitted and presented. 
To be introduced, evidence must satisfy an extensive list of rules defined in each jurisdiction.\footnote{In United States federal court, the rules governing these topics are the Federal Rules of Evidence. Each state court system has its own set of rules, but they are largely modeled after the federal rules, usually only deviating slightly~\cite{longlivefres}. Each country's court system has its own rules regarding evidence as well, but for simplicity, only the federal U.S. rules are discussed here.} 
In the US system, evidence must be both \emph{authenticated}\footnote{\textsc{Fed. R. Evid.} 901(a). Here, \emph{authentication} is a legal term of art, whose meaning differs completely from the cryptographic term of art in Section~\ref{sec:deniability}.} and \emph{relevant}\footnote{\textsc{Fed. R. Evid.} 104(b).} to be presented in court;
the judge decides whether a piece of evidence is satisfies these requirements. When helpful, \emph{expert testimony} may be brought in. We elaborate on these concepts below.

Finally, it is important that the legal system provides parties recourse against actual forgeries that others may attempt to use against them in court.\footnote{We are not aware of specific instances where forgeries of secure communication transcripts have come up as evidence in courts. We would be interested to hear of any examples. The closest we have come across is an interesting (as far as we know) hypothetical concern expressed by a Kenyan attorney that the blue ``double checkmark'' indicating that a WhatsApp message has been received and read might be misused as a ``proof'' in court that a recipient read a message, even if that message was deliberately deleted by the sender immediately afterward to prevent the recipient from reading it \cite[\S3.2.1]{MunyendoOSWAKR25}.} We do not believe the permissiveness of the evidentiary system 
generally means that parties are likely to be without recourse against actual forgeries --- at least not any more for deniable communications than for other types of forged evidence such as false verbal testimony, handwritten letters, doctored images, or cryptographically signed messages forged by password hacking. If one party forges a deniable communication, in practice, the other is likely to realize\footnote{An opposing party might not be able to identify a discrepancy if the evidence was in the sole custody of their opponent, such as a business record. But then deniability is not needed for a forgery, so deniability still does not introduce any novel challenges.}  ---
often even without any technical expertise. (``Wait, I never received that message!'') Essentially, the courts are designed to preempted this kind of forgery through their adversarial nature.

The system is far from perfect. But we believe, at least, that the same principles that underlie the current court system's imperfect efforts to address forgery of evidence apply and generalize well to deniable communication.

\subsection{Authentication}
\emph{Authentication} in court does not verify that a piece of evidence is true or ``authentic'' in the sense understood in colloquial speech or in computer science. It means that the party presenting the evidence must also produce evidence sufficient [that a jury could reasonably find by a preponderance of the evidence] that the item is what the proponent claims it is.\footnote{\textsc{Fed. R. Evid.} 901(a).} 
A screenshot of a chat transcript, for example, would most often be authenticated by the witness whose phone it is taken from testifying: ``That is a transcript of a conversation between me and the defendant. We had that conversation in Signal on Friday at 9 P.M.''\footnote{\textit{See} \textsc{Fed. R. Evid.} 901(b)(1).}

This may seem incredibly insufficient for those accustomed to cryptographic definitions of authentication.\footnote{When we write ``authentication,'' we mean authentication of evidence, not cryptographic authentication, unless otherwise specified.} However, the goals of cryptographic authentication and authentication of evidence in court are different; the latter is not weaker, but incomparable in its goals. The authentication process for evidence is not designed to determine that a piece of evidence is certain to be what the party introducing it says it is. It is designed to allow the parties to introduce their evidence while creating an opportunity for their opponent to challenge it.\footnote{\emph{See also} \cite[pp. 100--01]{Blanchette} for related discussion and historical context about authenticating digital evidence (in France, but the principles are similar).} Authentication via witness facilitates the assessment of whether evidence is credible based on the trustworthiness of the \textit{witness}, not just the \textit{evidence itself}. The witness is incentivized against lying because there are criminal penalties (for the crime of perjury) for doing so.\footnote{See, for example, 18 U.S.C. § 1621 (1994).} If the witness lies anyway, that may be indicated by other clues or contradictory evidence. This process works the same whether the evidence is a cryptographically deniable chat transcript, an unsigned letter, or a burnt piece of toast. The opposing party has a chance to cross-examine the witness in front of the jury. If there is reason to believe that the witness used deniability to alter the messages presented, they can ask the witness about it. The jury then decides how much bearing, if any, the transcript has on their verdict. Thus, the idea is that authentication of a deniable chat transcript via witness could be judged by the jury similarly to a witness describing a conversation held during a walk in the woods.

Alternatively, data copied during litigation directly from a person's device by a certified expert can be introduced without witness authentication; it is considered ``self-authenticating.''\footnote{\textit{See} \textsc{Fed. R. Evid.} 902(14).} Here, the copying party must provide notice to the non-copying party. The non-copying party has a chance to respond, and could raise the kinds of challenges discussed above.

\subsection{Relevance}

A piece of evidence is considered \emph{relevant} when it ``has a tendency to make a fact more or less probable than it would be without the evidence''\footnote{This phrasing may rankle the frequentists among us. Indeed, the Federal Rules of Evidence implicitly endorse Bayesian over frequentist probabilistic reasoning. We believe this is the right choice for the context; a full discussion is beyond the scope of this paper.} and the fact is ``of consequence'' to the case.\footnote{\textsc{Fed. R. Evid} 401.} ``Tendency to make a fact more or less probable'' is a flexible and inclusive phrasing that tries to get at the question of whether a person might reasonably assess a fact's likelihood of being true differently if they didn't have the evidence. The system is designed to put as much potentially useful evidence in front of the jury as possible. 

From a pure mathematical understanding of deniability, introducing a transcript from a communications service which implements deniability would not seem to have a tendency to make any fact more or less probable. After all, such a transcript is not cryptographic evidence that any party to the communication actually sent the pictured messages. What is more, within the cryptographic model, the deniability of the communication scheme may mean that \emph{any} proffered message is as likely as any other one to be the real one sent. By that same reasoning, introducing verbal testimony from a witness to a crime would not seem to have a tendency to make any fact more or less probable, because people can lie. 
However, when taking into account additional contextual factors, such as those introduced in our credibility framework, both the verbal testimony and the WhatsApp transcript could reasonably influence a juror's assessment of what likely happened --- and accordingly, both would generally meet the relevance standard.

The rules of evidence were written to determine the relevance of evidence before cryptographic verification existed, and aim
to handle, as fairly as possible, the scenarios in which it is most difficult to determine original from forgery. Cryptographic verification is not irrelevant to this process, and can streamline it when applicable --- but the broader process still exists to handle the vast majority of evidentiary situations where cryptographic verification is not determinative, including situations involving deniable communication. As there will almost always be factors outside of cryptographic models that indicate whether a transcript is the original or not, even judges who deeply understood cryptographic deniability would be justified in determining that a deniable chat transcript, in the context of broader circumstances and other evidence, could have at least a \textit{tendency} to make a fact more or less probable.

\subsection{Expert testimony}
Where a technical or scientific concept that is not widely understood is relevant to the case, courts also have mechanisms regarding \emph{expert testimony} to explain that concept.
Thus, a cryptographer by trade or education would  be able to testify as to the effects of deniability on the veracity of chat transcripts. However, it is worth noting that cases often suffer from ``dueling experts'' where both sides call expert witnesses with relevant qualifications who present apparently opposite information, even though there is only one side well supported by scientific evidence~\cite{duelingexperts}. 
Because judges and juries are not experts in the topics covered by most of the cases they try, they may side with the expert presenting bad evidence~\cite{duelingexperts}. Law enforcement agencies have a history of introducing dubious forensic evidence, including ballistic and fingerprint analysis, that is scientifically unsound, sometimes while attacking defendants' scientific experts inside and outside of the courtroom~\cite{lawenforcementexperts}. Similarly, for decades the tobacco industry avoided liability by presenting experts who dismissed the epidemiologic association between smoking and lung cancer because no one knew how cells actually became cancerous~\cite{smokingexperts}. Thus, bringing in experts is no panacea; it is important to consider the possibility that if a party tried to combat the presentation of forged deniable transcripts by presenting expert testimony, their opposing party might find an expert who would find some way to argue that the transcript was in fact the original, even though deniability is a settled cryptographic method. (To our knowledge, no such example has yet occurred.) Simply pointing out a discrepancy in a transcript without presenting expert testimony may sometimes be the best way to call the jury's attention to a possible forgery, without starting a battle of experts. That said, in some situations, such as where forgery by a sophisticated actor leveraging cryptographic deniability is suspected, bringing in an expert may be necessary to thoroughly explain the situation in the courtroom.

\subsection{Example}
\begin{figure}[ht]
    \centering
    \includegraphics[width=1\linewidth]{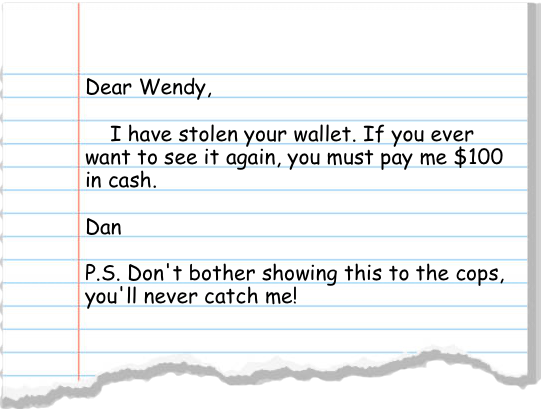} 
    \caption{Dan's ransom note to Wendy}
    \Description[Dan's ransom note to Wendy]{It reads: Dear Wendy, I have stolen your wallet. If you ever want to see it again, you must pay me \$100 in cash. Dan. P.S. Don't bother showing this to the cops, you'll never catch me!}
    \label{fig:ransom}
\end{figure}

To walk through the introduction of evidence at trial, consider a basic example: Prosecutor Pat is litigating the trial of Defendant Dan. Pat is claiming that Dan stole Witness Wendy's wallet and admitted to it in a handwritten ransom letter. Wendy approached Pat with the letter, shown in Figure~\ref{fig:ransom}, which requests \$100 in exchange for the wallet. Pat wants to introduce that letter as evidence at trial.

To introduce the letter, Pat has to show it to Dan's lawyer, authenticate it through a witness, in this case, Wendy, and ask the judge to enter it into evidence. To block Pat from introducing the evidence, Dan's lawyer can argue that it is insufficiently authenticated, or that it is not relevant and/or admissible. Unless Wendy comes off as clearly lying, the judge will allow Pat to present it to the jury. The jury will then have to decide how credible they think the letter is.

Now, imagine instead of giving Pat a handwritten letter, Wendy shows Pat a transcript from Signal. To introduce the transcript at trial through witness authentication, Pat will follow the exact same steps as the letter. Dan's lawyer can still argue that it is insufficiently authenticated, or that it is not relevant and/or admissible. And again, unless Wendy comes off as clearly lying, the judge will allow Pat to present it to the jury, which will decide how credible they think the transcript is.

Alternatively, Pat could introduce the transcript as self-authenticating if an expert copied it from Wendy's device. Then, Pat would have to provide notice to Dan's lawyer, who could perform additional checks such as comparing it to the transcript on Dan's phone. 

With either form of authentication, there are few scenarios where it would be worth the time of Dan's lawyers to argue against the introduction of the Signal transcript. The most likely would be if Pat tried to introduce a transcript copied from Wendy's phone that did not match the transcript on Dan's phone. It is important to note that this discrepancy would be immediately apparent to Dan's lawyers and easily impeachable even without technical knowledge of deniability. The mere fact of the discrepancy would enable Dan's lawyers to argue against the introduction of the evidence. While they could introduce technical experts to explain deniability, even without expert testimony, the jury's consideration of the evidence would be similar to consideration of conflicting testimony about a walk in the woods.

\section{Discussion}
\label{sec:discussion}

\paragraph{Legal authentication implicitly takes credibility into account.} By centering the human element, the courts' historical approach to authentication integrates credibility, and has the flexibility to adapt to different formats of evidence that are more or less credible --- that is, more or less easy to forge, more or less long-lived by default, and so on --- by taking their credibility into account in context. 
The approach of admitting evidence to court even if it is in a form that could easily be falsified is a deliberate choice to err on the side of allowing the fact finder (jury or judge) to assess all available information \emph{and its possible unreliability}, rather than strictly filtering what information they may consider; this deliberate choice was made before deniable communication technologies were available, but in a context where many other, more straightforwardly falsifiable forms of evidence were commonplace.

\paragraph{Deniability \emph{is effective} in courts today.} As such, our view is that deniability is serving its purpose effectively as it is treated in courts today---precisely because it forces reliance on traditional evidentiary procedures used for other forms of easily falsifiable evidence (e.g., corroborating witnesses), and precludes the use of cryptographic artifacts to bolster the case for the authenticity of some evidence. Our position diverges from prior work which has characterized courts unquestioningly admitting deniable communication transcripts into evidence as deniability being ``ineffective'' or ``irrelevant'' in courtroom settings \cite{yadav,collins}.

\paragraph{Existing legal systems are deeply imperfect.}
This paper is not intended to extol courts or evidentiary systems. Our points are conditioned on a legal system existing in its current form (and we believe many of our points are likely to generalize to a range of other legal systems). We argue that legal systems generally handle deniable communication evidence as their current evidentiary rules intend, rather than mishandling it due to a lack of technical expertise; and moreover, that those evidentiary rules have been developed with a detailed and explicit consideration of exactly the kinds of issues that cryptographic deniability raises, over a long time.
As such, the legal system does not require a paradigm shift to deal with deniability; rather, to the extent that the legal system does not deal with deniability satisfactorily, deniability is an additional example of the problems that the system has grappled with and will continue to face.

\paragraph{Judges' and laypeople's understanding of communication technologies does matter.}
Although, as we have discussed, all types of evidence are in some sense forgeable, judges' and jurors' perception of the fact and ease of forgeability varies between types of evidence and can have a significant bearing on the outcome of a case. For example, when hearing oral testimony, a jury will intuitively consider the fact that people can lie, but they may not as readily think of forgeability when weighing a chat transcript. A jury will also, in most cases, have an intuitive way of evaluating whether a witness is lying but likely will not have the same skills to evaluate digital evidence~\cite{rajendran}. As noted above, the evidentiary system offers some ``bootstrapping'' of this gap in knowledge by tying the authenticity of evidence back to oral testimony. 

The intuition of the average person, and therefore the average juror, regarding forgeability naturally changes over time. When photo editing software was first introduced, similar concerns were raised about whether juries could trust photographic evidence. Now, we expect that many jurors would intuitively understand that photographs can be manipulated (even if not exactly how), and to evaluate evidence with that knowledge even without prompting. Deepfakes, discussed in the following paragraph, are the newest technology to present similar challenges. Explaining the difficulty and likelihood of forgery to jurors in an intuitive way can be outcome determinative in court cases, so it is important for the administration of justice to better understand how people receive and interpret this kind of instruction. The studies discussed in Section~\ref{sec:related} are important initial steps in this process.

\paragraph{Deepfakes raise interesting new issues.} 
Deepfakes make the creation of believable forged images and videos far easier and more scalable than prior technologies.
Although the process of creating a deepfake is technically sophisticated under the hood, access to forgeries is quite easy at low cost for laypeople because users can create deepfakes via simple natural-language requests and, at least until recently, many online platforms incentivized such uses (e.g.,~\cite{mrdeepfakes}). The threshold of believability is also interesting --- because of media hype and political relevance, much of the public has heard about deepfakes even if they do not understand the underlying technology. In the courtroom, a general understanding of how the technology works may be relevant to decide when to allow AI-generated content in court proceedings, to prevent misunderstandings about any such content, and to detect surreptitious uses.\footnote{Such as~\cite{victimdeepfake}, a case involving a victim impact statement that appeared to be a video featuring a man killed in a road rage incident.).}

At least one case has already been thrown out of court, where plaintiffs tried to pass off deepfaked videos depicting real people as real videos of those people.\footnote{\textit{Mendones v. Cushman \& Wakefield, Inc.}, No. 23-CV-028772 (Cal. Super. Ct. Sept. 9, 2025).} In another case, a judge explicitly permitted a deepfaked victim impact statement that appeared to be a video featuring a man killed in a road rage incident talking about his own death, addressing the court and his killer~\cite{victimdeepfake}.\footnote{\textit{State v. Horcasitas}, No. CR2021-142720-001 (Az. Super. Ct. May 1, 2025).}

Deepfakes are likely to present significant challenges in the future. Courtrooms may not be the most important venue for this concern. 
Because of the high level of scrutiny and the likelihood that forensics will be applied to evidence in a legal case, a deepfake presented as an authentic image or video would arguably be fairly likely to be discovered, compared to deepfakes in many other contexts. That said, detection will get harder as technology advances.
We expect deepfakes to raise high-profile issues in and out of courtrooms in the coming years,\footnote{For example, President Trump has suggested that, in the future, ``If something happens that’s really bad, maybe I’ll have to just blame AI.''~\cite{trumpai}} and anticipate future research and public discourse on these important issues with interest. 

\section{Conclusion}
\label{sec:conclusion}

We see the value of deniability as lying largely outside the courts. 
But deniability can help provide users with communications secured against a variety of real threats, which support freedom of expression in practical scenarios.
Credibility is an essential consideration for implementing deniability features in practice,
and plays a major and previously underacknowledged role in the utility of deniable communications. 
Since credibility is inherently sociotechnical, we believe there is a great breadth of possible systems which could integrate our conceptual framework to provide novel value.

\begin{acks}
    We are grateful to Alastair Beresford, Joseph Bonneau, Ran Canetti, Daniel Collins, Simone Colombo, James Grimmelmann, Lo\"{\i}s Huguenin{-}Dumittan, Alice Hutchings, and our anonymous CSLAW 2026 reviewers for discussions and feedback.
\end{acks}

\printbibliography

@misc{aafs-forensics,
    author = {{American Academy of Forensic Sciences}},
    title = {What is forensic science?},
    url = {https://www.aafs.org/careers-forensic-science/what-forensic-science},
    note = {\textsc{permalink:} \url{https://web.archive.org/web/20250730015157/https://www.aafs.org/careers-forensic-science/what-forensic-science}}
}

@misc{aclu-disappearing,
    author = {Daniel Kahn Gillmor},
    title = {Disappearing Messages Don't Work --- And They’re Great},
    howpublished = {ACLU News \& Commentary},
    url = {https://www.aclu.org/news/privacy-technology/disappearing-messages-dont-work-and-theyre-great},
    year = {2025},
    note = {\textsc{permalink:} \url{https://perma.cc/FB9U-F5SN}}
}

@article{AG05,
    author       = {Alessandro Acquisti and
                  Jens Grossklags},
    title        = {Privacy and Rationality in Individual Decision Making},
    journal      = {{IEEE} Secur. Priv.},
    volume       = {3},
    number       = {1},
    pages        = {26--33},
    year         = {2005},
    url          = {https://doi.org/10.1109/MSP.2005.22},
    doi          = {10.1109/MSP.2005.22},
    bibsource    = {dblp computer science bibliography, https://dblp.org}
}

@book{Blanchette,
    author = {Jean-François Blanchette},
    title = {Burdens of Proof: Cryptographic Culture and Evidence Law in the Age of Electronic Documents},
    publisher = {MIT Press},
    year = {2012},
    isbn = {9780262017510}
}

@article{collins,
  author       = {Daniel Collins and
                  Simone Colombo and
                  Lo{\"{\i}}s Huguenin{-}Dumittan},
  title        = {Real-World Deniability in Messaging},
  journal      = {Proc. Priv. Enhancing Technol.},
  volume       = {2025},
  number       = {1},
  pages        = {320--340},
  year         = {2025}
}

@article{compelleddecryption,
    author = {Aloni Cohen and Sunoo Park},
    title = {Compelled Decryption and the Fifth Amendment: Exploring the Technical Boundaries},
    journal = {Harvard Journal of Law \& Technology},
    year = {2018},
    url = {https://jolt.law.harvard.edu/assets/articlePDFs/v32/32HarvJLTech169.pdf},
    note = {\textsc{permalink:} \url{https://perma.cc/RKT2-CX9G}}
}

@article{contextualintegrity,
    author = {Helen Nissenbaum},
    title = {Privacy as Contextual Integrity},
    journal = {Washington Law Review},
    volume = {79},
    issue = {1},
    year = {2004}
}

@InProceedings{deniableauthentication,
    author="Chakraborty, Suvradip and Hofheinz, Dennis and Maurer, Ueli and Rito, Guilherme",
    editor="Hazay, Carmit and Stam, Martijn",
    title="Deniable Authentication When Signing Keys Leak",
    booktitle="Advances in Cryptology -- EUROCRYPT 2023",
    year="2023",
    publisher="Springer Nature Switzerland",
    pages="69--100",
    isbn="978-3-031-30620-4",
    doi = {https://doi.org/10.1007/978-3-031-30620-4_3}
}

@misc{dictionarydeny,
    author = {The American Heritage Dictionary of the English Language},
    title = {Deniable},
    url = {https://ahdictionary.com/word/search.html?q=deniable},
    note = {\textsc{permalink:} \url{https://perma.cc/AQD9-QQRC}}
}

@misc{dictionaryforgery,
    author = {The American Heritage Dictionary of the English Language},
    title = {Forgery},
    url = {https://ahdictionary.com/word/search.html?q=forgery},
    note = {\textsc{permalink:} \url{https://perma.cc/XQK2-LWQX}}
}

@article{duelingexperts,
    author = {Kenton K. Yee},
    title = {Dueling Experts and Imperfect Verification},
    journal = {International Review of Law \& Economics},
    year = {2008},
    url = {https://papers.ssrn.com/sol3/papers.cfm?abstract_id=1312222}
}

@misc{eprint,
    author = {{International Association for Cryptologic Research}},
    title = {{Cryptology ePrint Archive}},
    url = {https://eprint.iacr.org}
}

@article{eyesbehavior,
    doi = {10.1371/journal.pone.0082055},
    author = {Bateson, Melissa AND Callow, Luke AND Holmes, Jessica R. AND Redmond Roche, Maximilian L. AND Nettle, Daniel},
    journal = {PLOS ONE},
    publisher = {Public Library of Science},
    title = {Do Images of ‘Watching Eyes’ Induce Behaviour That Is More Pro-Social or More Normative? A Field Experiment on Littering},
    year = {2013},
    url = {https://doi.org/10.1371/journal.pone.0082055},
}

@mastersthesis{fairoze-thesis,
    author = {Jaiden Fairoze},
    title = {On Zeroization in Secure Messaging},
    school = {University of Melbourne},
    year = {2021},
    note = {Unpublished}
}

@article{falsebottom,
    author={Ahmad, Shahzad and Rass, Stefan and Schartner, Peter},
    journal={IEEE Access}, 
    title={False-Bottom Encryption: Deniable Encryption From Secret Sharing}, 
    year={2023},
    volume={11},
}

@book{forensics-textbook1,
    author = {Casey, Eoghan},
    title = {Digital Evidence and Computer Crime: Forensic Science, Computers, and the Internet},
    year = {2011},
    isbn = {0123742684},
    publisher = {Academic Press, Inc.},
    edition = {3rd}
}

@book{forensics-textbook2,
    author = {Michael Hale Ligh and Andrew Case and Jamie Levy and Aaron Walters},
    title = {The Art of Memory Forensics: Detecting Malware and Threats in Windows, Linux, and Mac Memory},
    year = {2014},
    publisher = {Wiley},
    edition = {1st}
}

@misc{fullydeniable,
    author = {Ran Canetti and Sunoo Park and Oxana Poburinnaya},
    title = {Fully Deniable Interactive Encryption},
    howpublished = {Cryptology {ePrint} Archive, Paper 2018/1244},
    year = {2018},
    url = {https://eprint.iacr.org/2018/1244},
    note = {\textsc{permalink:} \url{https://perma.cc/JZV5-MWUL}}
}

@inproceedings{insecurekeys,
    author = {Li, Juanru and Lin, Zhiqiang and Caballero, Juan and Zhang, Yuanyuan and Gu, Dawu},
    title = {K-Hunt: Pinpointing Insecure Cryptographic Keys from Execution Traces},
    year = {2018},
    isbn = {9781450356930},
    publisher = {Association for Computing Machinery},
    url = {https://doi.org/10.1145/3243734.3243783},
    doi = {10.1145/3243734.3243783},
    booktitle = {Proceedings of the 2018 ACM SIGSAC Conference on Computer and Communications Security},
    pages = {412–425},
    numpages = {14},
    series = {CCS '18}
}

@book{knowncitizen,
    author = {Sarah E. Igo},
    title = {The Known Citizen},
    publisher = {Harvard University Press},
    year = {2018}
}

@article{lawenforcementexperts,
    author = {Paul C. Giannelli},
    title = {Daubert and Forensic Science: The Pitfalls of Law Enforcement Control of Scientific Research},
    journal = {University of Illinois Law Review},
    year = {2011},
    url = {https://illinoislawreview.org/wp-content/ilr-content/articles/2011/1/Giannelli.pdf},
    note = {\textsc{permalink:} \url{https://perma.cc/U9P4-HQLH}}
}

@article{longlivefres,
    author = {Daniel J. Capra and Liesa L. Richter},
    title = {Long Live the Federal Rules of Evidence!},
    journal = {George Mason Law Review},
    year = {2024},
    url = {https://lawreview.gmu.edu/wp-content/uploads/2023/12/Capra-Richter-31-Geo-Mason-L-Rev-1-2024.pdf},
    note = {\textsc{permalink:} \url{https://perma.cc/EQ4F-HVRC}}
}

@article{MC09,
    author = {Aleecia M McDonald and Lorrie Faith Cranor},
    title = {The cost of reading privacy policies},
    volume = {4},
    pages = {543--897},
    year = {2009},
    journal = {I/S: A Journal of Law \& Policy for the Information Society}
}

@article{mediumcomfort,
    title = {When is trust not enough? The role of perceived privacy of communication tools in comfort with self-disclosure},
    journal = {Computers in Human Behavior},
    volume = {26},
    number = {5},
    pages = {1120-1127},
    year = {2010},
    issn = {0747-5632},
    doi = {https://doi.org/10.1016/j.chb.2010.03.016},
    url = {https://www.sciencedirect.com/science/article/pii/S0747563210000567},
    author = {Nancy E. Frye and Michele M. Dornisch},
}

@article{mediummoderatesgroups,
    doi = {10.1371/journal.pone.0157827},
    author = {Lisiecka, Karolina AND Rychwalska, Agnieszka AND Samson, Katarzyna AND Łucznik, Klara AND Ziembowicz, Michał AND Szóstek, Agnieszka AND Nowak, Andrzej},
    journal = {PLOS ONE},
    publisher = {Public Library of Science},
    title = {Medium Moderates the Message. How Users Adjust Their Communication Trajectories to Different Media in Collaborative Task Solving},
    year = {2016},
    month = {06},
    volume = {11},
    url = {https://doi.org/10.1371/journal.pone.0157827},
}

@article{mediumshapesindividuals,
    author = {Oba, Demi and Berger, Jonah},
    title = {How communication mediums shape the message},
    journal = {Journal of Consumer Psychology},
    volume = {34},
    number = {3},
    doi = {https://doi.org/10.1002/jcpy.1372},
    url = {https://myscp.onlinelibrary.wiley.com/doi/abs/10.1002/jcpy.1372},
    year = {2024}
}

@misc{mrdeepfakes,
    author = {Emanuel Maiberg and Samantha Cole},
    title = {Mr. Deepfakes, the Biggest Deepfake Porn Site on the Internet, Says It’s Shutting Down for Good},
    url = {https://www.404media.co/mr-deepfakes-the-biggest-deepfake-porn-site-on-the-internet-says-its-shutting-down-for-good/},
    year = {2025},
    note = {\textsc{permalink:} \url{https://perma.cc/GE6L-DG2K}}
}

@inproceedings{MunyendoOSWAKR25,
  author       = {Collins W. Munyendo and
                  Kentrell Owens and
                  Faith Strong and
                  Shaoqi Wang and
                  Adam J. Aviv and
                  Tadayoshi Kohno and
                  Franziska Roesner},
  editor       = {Marina Blanton and
                  William Enck and
                  Cristina Nita{-}Rotaru},
  title        = {"You Have to Ignore the Dangers": User Perceptions of the Security
                  and Privacy Benefits of WhatsApp Mods},
  booktitle    = {{IEEE} Symposium on Security and Privacy, {SP} 2025, San Francisco,
                  CA, USA, May 12-15, 2025},
  pages        = {4515--4533},
  publisher    = {{IEEE}},
  year         = {2025},
  url          = {https://doi.org/10.1109/SP61157.2025.00087},
  doi          = {10.1109/SP61157.2025.00087},
  bibsource    = {dblp computer science bibliography, https://dblp.org}
}

@misc{nist-forensics,
    author = {{National Institute of Standards and Technology}},
    title = {What is forensic science?},
    url = {https://www.nist.gov/forensic-science},
    note = {\textsc{permalink:} https://perma.cc/4AUR-NUNL}
}

@misc{nyt-privacy-policies,
    author = {Kevin Litman-Navarro},
    title = {We Read 150 Privacy Policies. {They} Were an Incomprehensible Disaster.},
    journal = {The New York Times},
    year = {2019},
    howpublished = {\url{https://www.nytimes.com/interactive/2019/06/12/opinion/facebook-google-privacy-policies.html}},
    note = {\textsc{permalink:} \url{https://perma.cc/5XJJ-4CQV}}
}

@misc{ohchr,
    author = {Kaye, David},
    title = {Report on encryption, anonymity, and the human rights framework},
    journal = {Human Rights Council},
    year = {2015},
    howpublished = {\url{https://www.undocs.org/A/HRC/29/32}},
    note = {\textsc{permalink:} \url{https://perma.cc/QN7A-R9V5}}
}

@misc{originaldeniability,
    author = {Ran Canetti and Cynthia Dwork and Moni Naor and Rafi Ostrovsky},
    title = {Deniable Encryption},
    howpublished = {Cryptology {ePrint} Archive, Paper 1996/002},
    year = {1996},
    url = {https://eprint.iacr.org/1996/002},
    note = {\textsc{permalink:} \url{https://perma.cc/D5WL-4JNJ}}
}

@article{privacy-paradox-myth,
    author = {Daniel J. Solove},
    title = {The Myth of the Privacy Paradox},
    journal = {George Washington Law Review},
    volume = {89},
    issue = {1},
    year = {2021}
}

@InProceedings{quantum-non-retention,
    author="Bartusek, James and Goyal, Vipul and Khurana, Dakshita and Malavolta, Giulio and Raizes, Justin and Roberts, Bhaskar",
    editor="Joye, Marc and Leander, Gregor",
    title="Software with Certified Deletion",
    booktitle="Advances in Cryptology -- EUROCRYPT 2024",
    year="2024",
    publisher="Springer Nature Switzerland",
    pages="85--111",
    isbn="978-3-031-58737-5",
    doi = {https://doi.org/10.1007/978-3-031-58737-5_4}
}

@inproceedings{rajendran,
    author = {Rajendran, Anamika and Yadav, Tarun Kumar and Al-Jbour, Malek and Mares Solano, Francisco Manuel and Seamons, Kent and Reynolds, Joshua},
    title = {Deniable Encrypted Messaging: User Understanding after Hands-on Social Experience},
    year = {2024},
    isbn = {9798400717963},
    publisher = {Association for Computing Machinery},
    url = {https://doi.org/10.1145/3688459.3688479},
    doi = {10.1145/3688459.3688479},
    booktitle = {Proceedings of the 2024 European Symposium on Usable Security},
    pages = {155–171},
}

@inproceedings{reitinger,
    author={Nathan Reitinger and Nathan Malkin and Omer Akgul and Michelle L. Mazurek and Ian Miers},
    booktitle={2023 IEEE Symposium on Security and Privacy (SP)}, 
    title={Is Cryptographic Deniability Sufficientƒ Non-Expert Perceptions of Deniability in Secure Messaging}, 
    year={2023},
    pages={274-292},
    doi={10.1109/SP46215.2023.10179361}
}

@misc{signaldocs,
    author = {Moxie Marlinspike},
    title = {The X3DH Key Agreement Protocol},
    year = {2016},
    publisher = {Signal},
    url = {https://signal.org/docs/specifications/x3dh},
    note = {\textsc{permalink:} \url{https://perma.cc/LCV5-KW9T}}
}

@article{smokingexperts,
    author = {Steve C. Gold},
    title = {A Fitting Vision of Science for the Courtroom},
    journal = {Wake Forest Journal of Law and Policy},
    year = {2013},
    url = {https://ssrn.com/abstract=2101454}
}

@article{taleoftwo,
    author = {Herbert L. Packer},
    title = {A Tale of Two Typewriters},
    journal = {Stanford Law Review},
    year = {1958},
    url = {https://doi.org/10.2307/1226822}
}

@misc{teenagersinternet,
    author = {Katie Eichhorn},
    title = {Why an internet that nevers forgets is especially bad for young people},
    url = {https://www.technologyreview.com/2019/12/27/131123/internet-that-never-forgets-bad-for-young-people-online-permanence/},
    year = {2019},
    note = {\textsc{permalink:} \url{https://perma.cc/U4CP-PEDL}}
}

@misc{truecryptusage,
    author = {Brian Prince},
    title = {The Mystery Of The TrueCrypt Encryption Software Shutdown},
    howpublished = {Dark Reading},
    url = {https://www.darkreading.com/endpoint-security/the-mystery-of-the-truecrypt-encryption-software-shutdown},
    year = {2014},
    note = {\textsc{permalink:} \url{https://web.archive.org/web/20240910142704/https://www.darkreading.com/endpoint-security/the-mystery-of-the-truecrypt-encryption-software-shutdown}}
}

@article{trustnotenough,
    title = {When is trust not enough? The role of perceived privacy of communication tools in comfort with self-disclosure},
    journal = {Computers in Human Behavior},
    year = {2010},
    doi = {https://doi.org/10.1016/j.chb.2010.03.016},
    url = {https://www.sciencedirect.com/science/article/pii/S0747563210000567},
    author = {Nancy E. Frye and Michele M. Dornisch},
}

@misc{trumpai,
    author = {Michelle L. Price},
    title = {Trump says video showing items thrown from White House is AI after his team indicates it’s real},
    howpublished = {AP News},
    url = {https://apnews.com/article/trump-white-house-window-ai-9dad119bb6de1519582db036dc0726d7},
    year = {2025},
    note = {\textsc{permalink:} \url{https://web.archive.org/web/20250904144114/https://apnews.com/article/trump-white-house-window-ai-9dad119bb6de1519582db036dc0726d7}}
}

@inproceedings{vankleek2017,
    author = {Van Kleek, Max and Liccardi, Ilaria and Binns, Reuben and Zhao, Jun and Weitzner, Daniel J. and Shadbolt, Nigel},
    title = {Better the Devil You Know: Exposing the Data Sharing Practices of Smartphone Apps},
    year = {2017},
    booktitle = {Proceedings of the 2017 CHI Conference on Human Factors in Computing Systems},
}

@misc{vatandas,
    author = {Nihal Vatandas and Rosario Gennaro and Bertrand Ithurburn and Hugo Krawczyk},
    title = {On the Cryptographic Deniability of the Signal Protocol},
    howpublished = {Cryptology {ePrint} Archive, Paper 2021/642},
    year = {2021},
    url = {https://eprint.iacr.org/2021/642},
    note = {\textsc{permalink:} \url{https://perma.cc/UY62-9BZJ}}
}

@misc{victimdeepfake,
    author = {Matthew Gault and Jason Koebler},
    title = {'I Loved That AI:' Judge Moved by AI-Generated Avatar of Man Killed in Road Rage Incident},
    howpublished = {404 Media},
    year = {2025},
    url = {https://www.404media.co/i-loved-that-ai-judge-moved-by-ai-generated-avatar-of-man-killed-in-road-rage-incident/},
    note = {\textsc{permalink:} \url{https://perma.cc/6SX9-S46X}}
}

@inproceedings {yadav,
    author = {Tarun Kumar Yadav and Devashish Gosain and Kent Seamons},
    title = {Cryptographic Deniability: A Multi-perspective Study of User Perceptions and Expectations},
    booktitle = {32nd USENIX Security Symposium (USENIX Security 23)},
    year = {2023},
    isbn = {978-1-939133-37-3},
    pages = {3637-3654},
    url = {https://www.usenix.org/conference/usenixsecurity23/presentation/yadav},
    publisher = {USENIX Association},
    note = {\textsc{permalink:} \url{https://perma.cc/PDM5-LMSW}}
}

\end{document}